\documentclass[prb,twocolumn,showpacs,superscriptaddress]{revtex4}

\usepackage{graphicx}
\usepackage{amsmath}
\usepackage{amssymb}
\usepackage{epsfig}

\renewcommand{\v}[1]{{\bf #1}}

\def\eqa{\begin{eqnarray}}
\def\eea{\end{eqnarray}}
\newcommand{\eq}{\begin{equation}}
\newcommand{\ee}{\end{equation}}
\newcommand{\nn}{\nonumber\\}

\newcommand{\<}{\langle}
\renewcommand{\>}{\rangle}

\renewcommand{\Im}{{\rm Im}}
\renewcommand{\Re}{{\rm Re}}
\newcommand{\p}{\partial}

\newcommand{\ua}{\uparrow}
\newcommand{\da}{\downarrow}
\newcommand{\ra}{\rightarrow}

\newcommand{\al}{\alpha}
\newcommand{\bt}{\beta}
\newcommand{\del}{\delta}
\newcommand{\Del}{\Delta}
\newcommand{\eps}{\epsilon}

\newcommand{\ga}{\gamma}
\newcommand{\Ga}{\Gamma}

\newcommand{\la}{\lambda}
\newcommand{\La}{\Lambda}

\renewcommand{\th}{\theta}

\newcommand{\si}{\sigma}

\usepackage{color}

\begin{document}

\title{Chiral and helical $p$-wave superconductivity in doped bilayer BiH}

\author{Lin Yang}
\affiliation{National Laboratory of Solid State Microstructures \& School of Physics, Nanjing
	University, Nanjing, 210093, China}

\author{Wan-Sheng Wang}
\affiliation{Department of Physics, Ningbo University, Ningbo 315211, China}

\author{Da Wang}
\affiliation{National Laboratory of Solid State Microstructures \& School of Physics, Nanjing
	University, Nanjing, 210093, China}

\author{Qiang-Hua Wang}
\email{qhwang@nju.edu.cn}
\affiliation{National Laboratory of Solid State Microstructures \& School of Physics, Nanjing
	University, Nanjing, 210093, China}
\affiliation{Collaborative Innovation Center of Advanced Microstructures, Nanjing University, Nanjing 210093, China}

\begin{abstract}
We investigate the superconductivity (SC) driven by correlation effects in electron-doped bilayer BiH near a type-II van Hove singularity (vHS). By functional renormalization group, we find triplet $p$-wave pairing prevails in the interaction parameter space, except for spin density wave (SDW) closer to the vHS or when the interaction is too strong. Because of the large atomic spin-orbital coupling (SOC), the $p$-wave pairing occurs between equal-spin electrons, and is chiral and two-fold degenerate. The chiral state supports in-gap edge states, even though the low energy bands in the SC state are topologically trivial. The absence of mirror symmetry allows Rashba SOC that couples unequal spins, but we find its effect is of very high order, and can only drive the chiral $p$-wave into helical $p$-wave deep in the SC state. Interestingly, there is a six-fold degeneracy in the helical states, reflected by the relative phase angle $\th=n\pi/3$ (for integer $n$) between the spin components of the helical pairing function. The phase angle is shown to be stable in the vortex state. 
\end{abstract}

\pacs{71.27.+a, 74.20.-z, 74.20.Rp}
%

\maketitle


\section{Introduction}

There are much interest in the search for triplet superconductivity (SC), with the hope that a $p$-wave triplet superconductor may host Majorana zero modes in vortices \cite{Ivanov, Read} and hence could serve as the platform for topological quantum computing.\cite{Read, Nayak}
Usually, triplet pairing is related to spin fluctuations at small wavevector (or long wavelength). Such fluctuations are enhanced when the Fermi level is close to a van Hove singularity (vHS), or when Fermi pockets are close to each other in the momentum space. \cite{wangyao,frg3,frg7} However, if the vHS is on the zone boundary, as in the case of $\mathrm{Sr_2RuO_4}$, the vH momenta are time-reversal invariant up to a reciprocal vector. Such a vHS is classified as of type-I. \cite{yaohong} In this case, triplet Cooper pairing on opposite vH momenta is forbidden by Pauli exclusion. To get around this destructive effect, a recent proposal is to look for systems with type-II vHS, where vH momenta are not time-reversal invariant. \cite{yaohong} BC$_3$ is such a material, and theoretical study implies that $p$-wave pairing is likely near the vHS, \cite{yaohong,wangyao} although the transition temperature is sizable only if the local bare interaction is moderately large. \cite{wangyao} The other material that has a type-II vHS is the bilayer BiH. In the absence of doping, the material is a quantum spin-Hall insulator with a large indirect band gap. \cite{Song, Liu}. Electron doping drives the Fermi level toward the type-II vHS. Hatree-Fock mean field theory shows that triplet pairing could be stabilized even by small electron doping of the insulating compound, with a small electron pocket around the zone center and a Fermi level relatively far from the type-II vHS. \cite{Yang} However, the mean field coupling constant is  about $\la \sim 10^{-2}$, implying a tiny transition temperature according to $T_c\propto e^{-1/\la}$. This motivates us to consider filling levels near the type-II vHS (realizable by higher electron doping). With enhanced density of states (DOS) near the vHS, the system becomes more susceptible to instabilities under interactions. On the other hand, fluctuations in the  particle-particle (pp) and particle-hole (ph) channels are intertwined. To gain a better estimate of the transition temperature, it is necessary to treat the pp and ph channels on equal footing. For this purpose we resort to the singular-mode functional renormalization group (SM-FRG), which has been applied in various contexts, \cite{frg1, frg2, frg3, frg4, frg5, frg6, frg7, frg8, frg9} and is particularly useful in the presence of SOC, which is significant in BiH because of the heavy Bi element.

The main results are as follows. By SM-FRG we find local triplet $p$-wave pairing prevails in the interaction parameter space, except for spin density wave (SDW) closer to the vHS or when the interaction is too strong. The typical SC $T_c$ is $0.1\sim 1$ meV near the phase boundary between SC and SDW. The $p$-wave pairing is between equal-spin electrons, with a sign change between the two sublattices, and is chiral and two-fold degenerate. The chiral state supports in-gap edge states, but interestingly the two low energy bands in the SC state are topologically trivial. The Rashba SOC, from the absence of mirror symmetry, can couple unequal spins, but its effect is so weak that it can only drive the chiral $p$-wave into helical $p$-wave deep in the SC state. The helical states are six-fold degenerate in terms of the relative phase angle between the two components in the helical pairing function, and is shown to be stable even in the vortex state.

The rest of this paper is organized as follows. We specify the model in Sec.\ref{sec:model}, followed by the SM-FRG in Sec.\ref{sec:smfrg} to find the favorable pairing function and typical transition temperature. We then discuss in Sec.\ref{sec:chiral} the topological property of the chiral $p$-wave SC revealed by SM-FRG, and discuss the helical $p$-wave SC under a Rashba SOC in Sec.\ref{sec:helical}. Finally, Sec. \ref{sec:summary} contains a summary of this work and some perspective remarks.

\section{Model}\label{sec:model}

\begin{figure}
	\includegraphics[width=0.98\columnwidth]{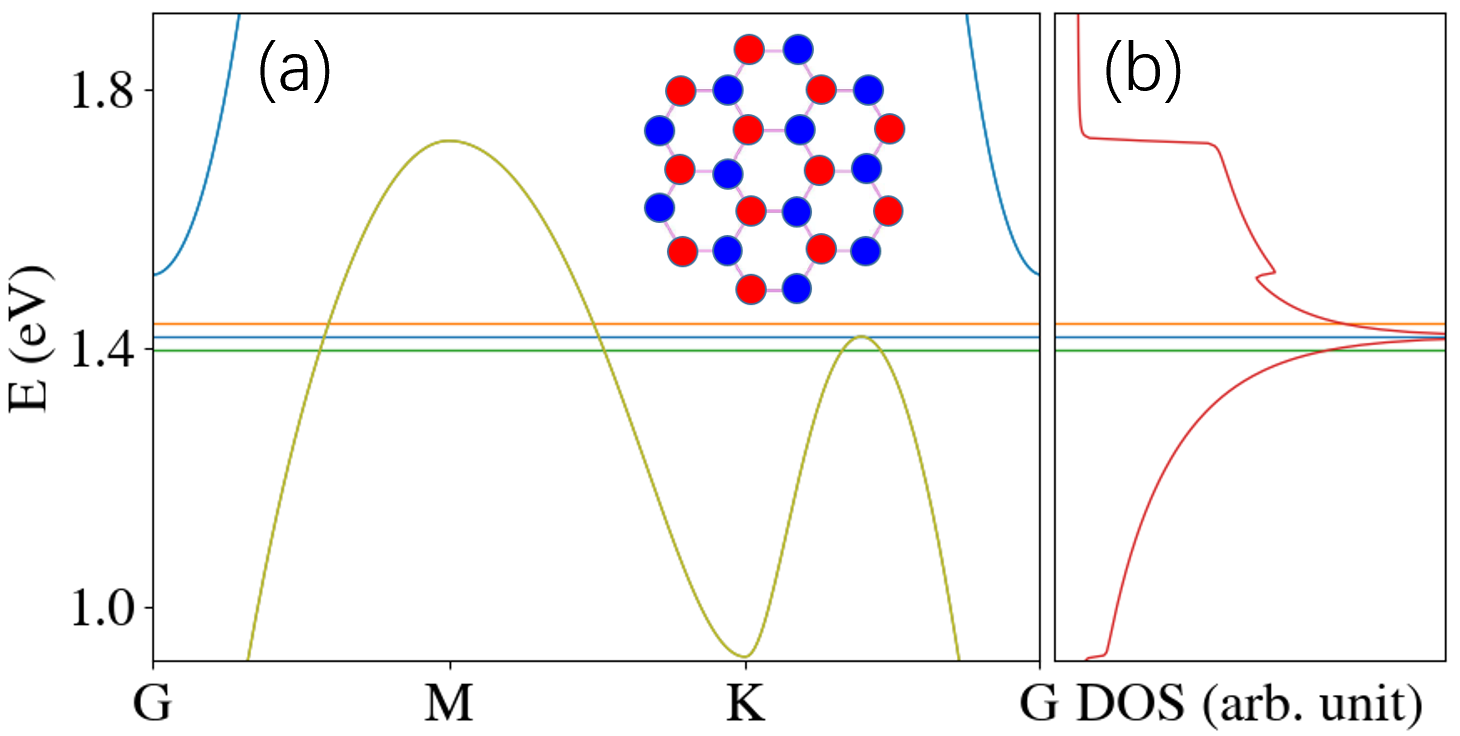}
	\caption{(a) The band dispersion near the vHS along high symmetry cuts. The horizontal lines highlight typical Fermi levels near the vHS under discussion, corresponding to the filling levels $n=0.5$, $0.55$ and $0.61$, respectively. The inset shows the buckled honeycomb lattice. The red (blue) circles present bismuth atoms slightly above (below) the central plane, hydrogenated from above (below). (b) The DOS near the vHS.}\label{fig:model}
\end{figure}

The structure of the bilayer BiH is shown in the inset of Fig.\ref{fig:model}(a). The bismuth atoms are located on a slightly buckled honeycomb lattice, with sublattice A and B hydrogenated from above and below, respectively. According to first principle calculation, \cite{Yang, Song, Liu} the band electrons near the Fermi level are mainly derived from the $x$- and $y$- orbitals of Bi. A corresponding tight-binding model can be written as \cite{Song, Liu, Yang, GFZhang},
\eqa H_0= &&\sum_{i\v b}  \psi^\dag_{i} ~t_\v b ~\psi_{i+\v b} -\sum_{i} \psi_i^\dag (\mu + \la\tau_2\si_3)\psi_i.\eea
Here $\mu$ is the chemical potential, and $\la$ the atomic spin-orbital coupling. The spinor $\psi_i=(c_{ix\ua}, c_{iy\ua}, c_{ix\da}, c_{iy\da})^t$, where $c_{i\al\si}$ annihilates an electron at site $i$ with orbital $\al\in (x, y)$ and spin $\si \in (\ua, \da)$. The Pauli matrices $\tau$ and $\si$ act in the orbital and spin basis, respectively. The vector $\v b$ (of both signs) connects two sites, associated with the hopping matrix (in orbital basis) defined by the elements
\eqa t_\v b^{\al\bt} = t_b^{s} (\v n_\al\cdot \hat{b}) (\v n_\bt\cdot\hat{b}) + t_b^{p} (\v n_\al\times\hat{b})\cdot (\v n_\bt\times \hat{b}),\eea
where $\hat{b}=\v b/b$ and $\v n_{\al}=\hat{x}\del_{\al x}+\hat{y}\del_{\al y}$ are unit vectors, and $t_b^{s/p}$ are Koster-Slater coefficients \cite{Slater} for bonds of a given length $b$. Following Ref.\cite{Yang}, we take $t_{1}^{s} = 1.79$ eV, and
\eqa (t_{1}^{p}, t_2^{s}, t_2^{p},\la)=(-0.45, 0.04, -0.15, 0.35) t_1^{s},\eea  where the subscripts 1 and 2 refer to the first and second neighbor bonds, respectively. The band dispersion near the vHS along high-symmetry cuts is shown in the main panel of Fig.\ref{fig:model}(a), and the corresponding DOS is shown in (b).  Notice that each band is two-fold degenerate by time-reversal (TR) invariance of $H_0$ and the local nature of the atomic SOC. It is seen that the vHS is located between $G$ and $K$, hence is of type-II.  Notice the material has $S_6$ symmetry (about the center of the holo hexagon), but in the above two-dimensional (2d) model, inversion alone is also a symmetry. This symmetry makes parity a good quantum number, and dictates $(p_x,p_y)$ to be a doublet representation.

We consider the following local interactions,
\eqa H_I =&&\sum_{i\al} U n_{i\al\ua} n_{i\al\da} + \sum_{i,\al>\bt} U' n_{i\al}n_{i\bt} \nn
   + && \sum_{i,\al>\bt,\si,\si'} J_H c_{i\al\si}^\dag c_{i\bt\si} c_{i\bt\si'}^\dag c_{i\al\si'}\nn
   + && \sum_{i,\al \bt} J_P c_{i\al\ua}^\dag c_{i\al\da}^\dag c_{i\bt\da}c_{i\bt\ua} + \sum_{\<ij\>\in {\rm NN}} V n_i n_j. \eea
Here $n_{i\al\si}=c_{i\al\si}^\dag c_{i\al\si}$, $n_{i\al}=\sum_\si n_{i\al\si}$, $n_i=\sum_\al n_{i\al}$,  $U$ ($U'$) is the intra- (inter-) orbital interaction, $J_H$ ($J_P$) is the Hund's rule coupling (pair hopping), and finally $V$ is the Coulomb interaction on nearest-neighbor bonds. We use the Kanomori relations $U=U'+2J_H$ and $J_H=J_P$ to reduce the number of independent interaction parameters. Furthermore, $J_H/U\sim 0.4$ according to the first principle calculation.\cite{Yang}

\begin{figure}
	\includegraphics[width=0.9\columnwidth]{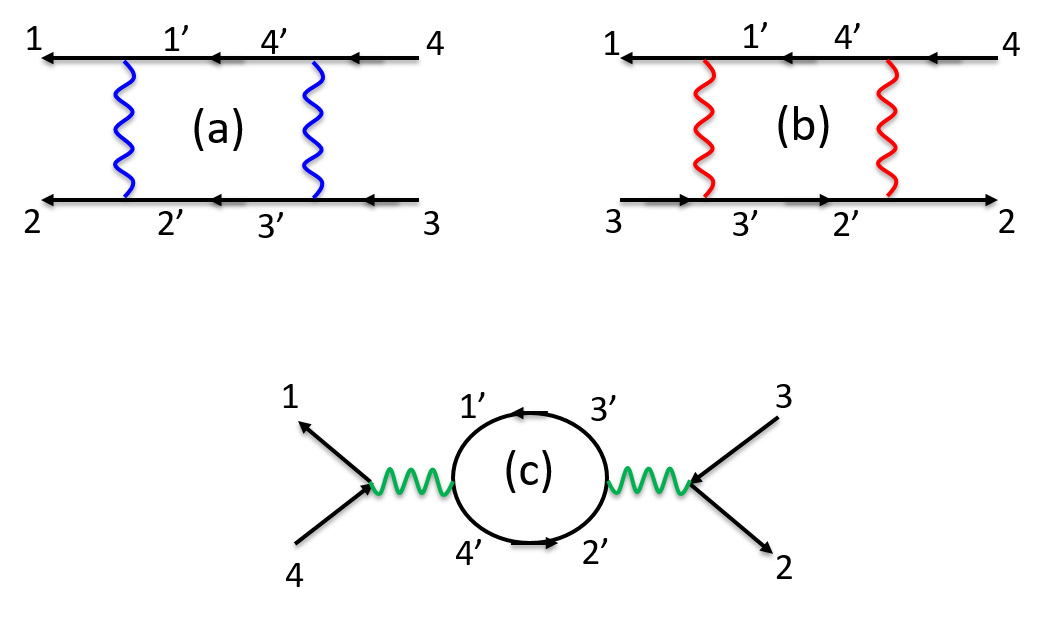}
	\caption{One-loop diagrams contributing to $\p\Ga_{1234}/\p\La$, quadratic in $\Ga$ itself (wavy lines, fully antisymmetrized with respect to incoming or outgoing fermions, labelled by the numerical indices). The color of the wavy line signifies scattering of fermion bilinears in the (a) pairing, (b) crossing and (c) direct channel.}\label{fig:feynman}
\end{figure}

\section{SM-FRG}\label{sec:smfrg}
The interactions can lead to competing collective fluctuations in particle-hole (ph) and particle-particle (pp) channels, which we handle by SM-FRG. The idea of FRG \cite{Wetterich} is to obtain the one-particle-irreducible 4-point interaction vertices $\Ga_{1234}$ (where numerical index labels single-particle state) for quasi-particles above a running infrared energy cut off $\La$ (which we take as the lower limit of the continuous Matsubara frequency). Starting from $\La=\infty$ where $\Ga$ is specified by the bare parameters in $H_I$, the contribution to the flow (toward decreasing $\La$) of the vertex, $\p\Ga_{1234}/\p\La$, is illustrated in Fig.\ref{fig:feynman}. At each stage of the flow, we decompose $\Ga$ in terms of eigen scattering modes (separately) in the pp and ph channels to find the negative leading eigenvalue (NLE) at each collective momentum $\v q$. The divergence of the most negative eigenvalue (MNE) at a scale $\La_c$ signals an emerging order at a transition temperature $T_c\sim  \La_c$, \cite{note} with the internal microscopic structure described by the eigenfunction. The technical details can be found elsewhere,\cite{frg1, frg2, frg3, frg4, frg5, frg6} and also in the self-contented SM.\cite{SM}

\begin{figure}
	\includegraphics[width=0.8\columnwidth,clip]{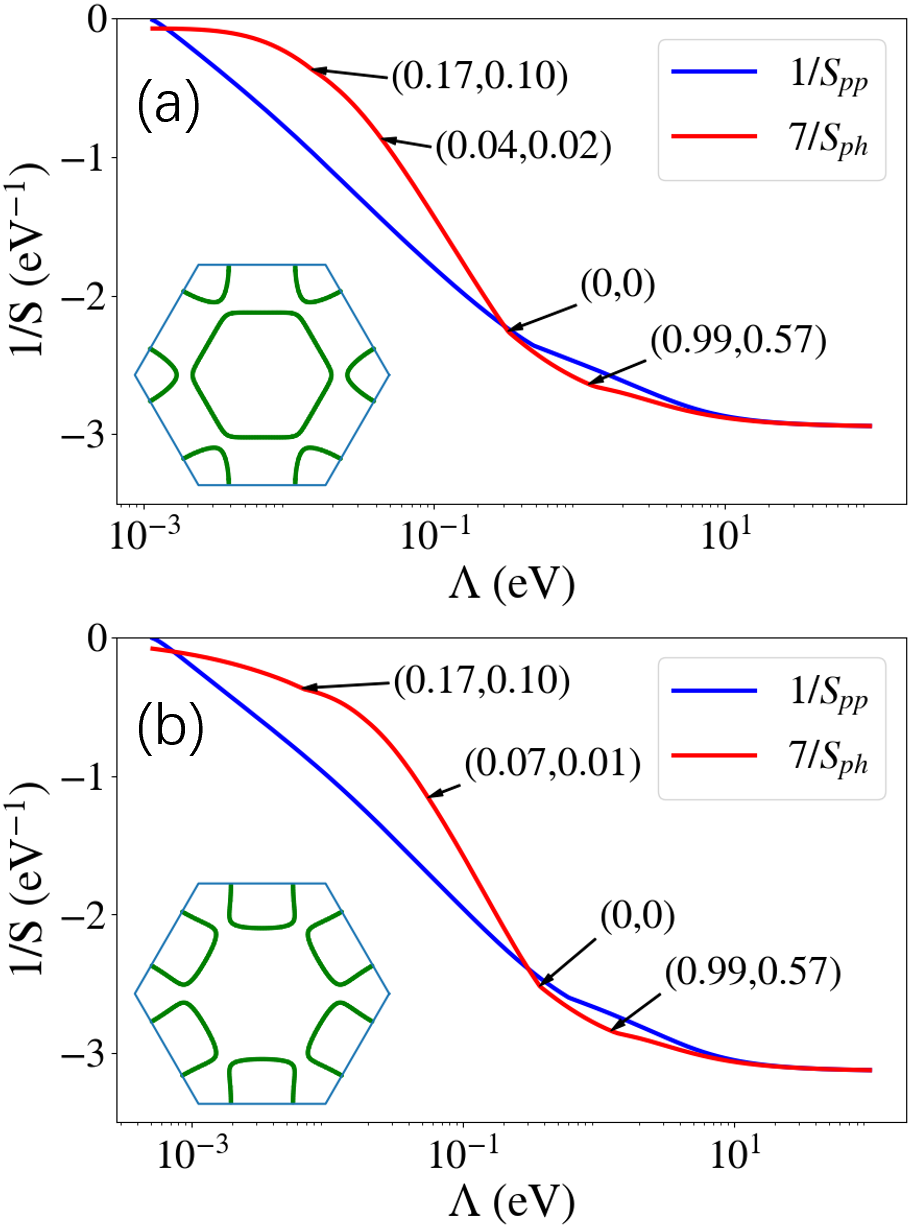}
	\caption{Flow of the MNE's $S_{\rm pp,ph}$, plot as $1/S_{\rm pp,ph}$ for clarity, as $\La$ is decreased for (a) $n=0.50$, $U=1.7$ eV, $J_H=0.4U$ and $V=0$, and (b) $n=0.61$, $U=1.6$ eV, $J_H=0.4U$ and $V=0$. The arrows are snapshots of $\v Q/\pi$ (with symmetric images in momentum space) in the ph channel. The insets show the Fermi pockets.}\label{fig:frg}
\end{figure}

First, we consider an electron density $n=0.50$. The Fermi level corresponds to the lowest horizontal line in Fig.\ref{fig:model}(a). The Fermi pockets concentrate on $G$ and $K$ points, see the inset of Fig.\ref{fig:frg}(a). The main panel shows the FRG flow of the MNE's $S_{\rm pp, ph}$ in the pp and ph channels for $U=1.7$ eV, $J_H=0.4U$ and $V=0$. The arrows snapshot the collective momentum $\v Q$ associated with the MNE in the ph channel. From the eigenfunction we find the MNE scattering mode is local in real space and spin-like, with moment along $z$. (The anisotropy in the spin moment is caused by the large atomic SOC.) Therefore, the system develops strong small-momentum spin fluctuations at low energy scales, and this can be ascribed to the scattering between the $G$ and $K$ pockets, separated by the vHS. However, $S_{\rm ph}$ saturates eventually because the Fermi level is not exactly at the vHS so that the phase space diminishes for low-energy ph excitations at a particular momentum. On the other hand, the MNE in the pp channel is initially weak, and is enhanced along with the ph channel. In the final stage, the pp channel diverges on its own at $\La_c=1.16\times 10^{-3}$ eV, implying SC order. We find the pairing function (given by the eigenfunction of the MNE scattering mode) is two-fold degenerate, $\phi(\v k)\sim i\tau_2 s_3 (\si_1\pm i\si_2)i\sigma_2$ (up to a global scale), where $s_3$ is the Pauli matrix in the sublattice basis. This describes local orbital-singlet (via $i\tau_2$) and equal-spin triplet pairing, with a sign change on the two sublattices (via $s_3$). Since both $\tau_2$ and $s_3$ transform as $f$-wave under rotation (about the center of a holo hexagon) in the 2d model, the entire pairing function transforms as $p_x\pm i p_y$ under combined rotations of spin, orbital and lattice. The obtained pairing function is similar to that in Ref.\cite{Yang} where the Fermi level is far below the vHS, suggesting the robustness of this type of pairing symmetry. Indeed, similar results are obtained for $n=0.61$, corresponding to the top line in Fig.\ref{fig:model}(a), with $U=1.6$ eV, $J_H=0.4U$ and $V=0$, although the Fermi level is slightly above the vHS and the Fermi pockets now concentrate on the M points, see the inset in Fig.\ref{fig:frg}(b).

\begin{figure}
	\includegraphics[width=0.8\columnwidth,clip]{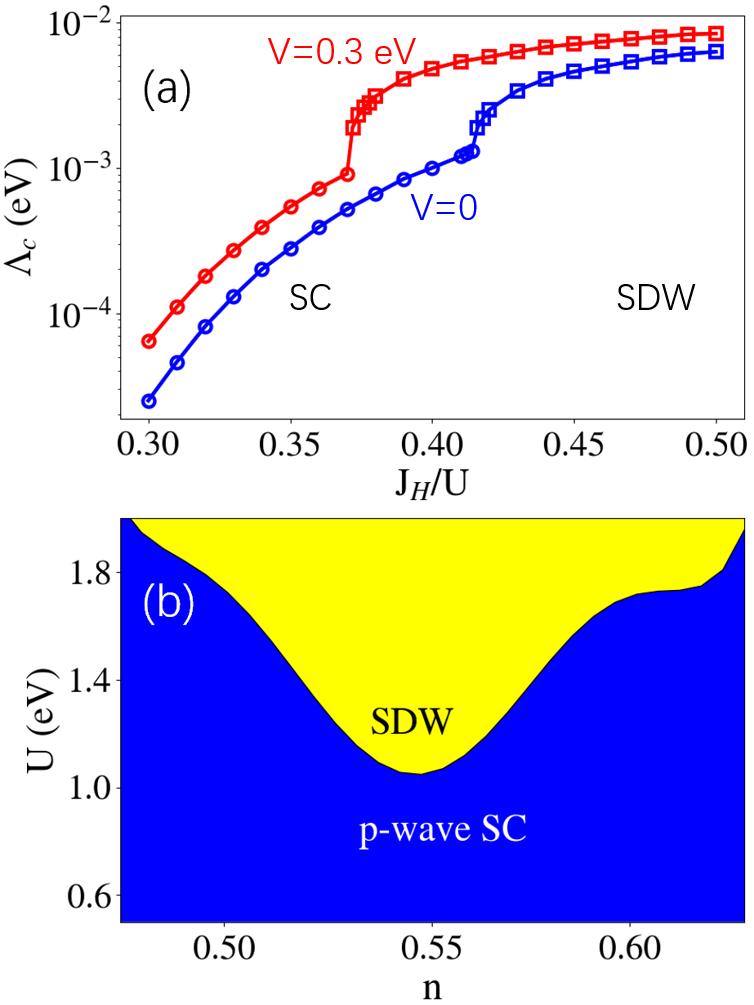}
	\caption{(a) The divergence scale $\La_c$ as a function of $J_H$ for $U=1.7$ eV. The open circles (squares) are for the SC (SDW) phase, and the blue (red) color is for $V=0$ ($V=0.3$ eV). Lines are drawn to guide the eye. (b) Schematic phase diagram in the parameter space with $J_H=0.4 U$ and $V=0$. The vH filling is $n= 0.55$.}\label{fig:phase}
\end{figure}

Figure \ref{fig:phase}(a) shows the divergence scale $\La_c$ increases quasi exponentially as a function of $J_H$, for $U=1.7$ eV. The ordered state is $p$-wave SC (open circles) at lower values of $J_H$, and the SDW state (open squares) if $J_H$ is increased further. We should point out that in the Hatree-Fock mean field theory \cite{Yang} the SC state requires $J_H\geq U/3$, but this is not necessary in our FRG since the the pp and ph channels are naturally intertwined, although a larger $J_H$ does help SC before the SDW order sets in. On the other hand, the results at $V=0$ (blue lines) and $V=0.3$ eV (red lines) show that a small nearest-neighbor Coulomb interaction can enhance $\La_c$ in both phases.
Fig.\ref{fig:phase}(b) is a schematic phase diagram in the $(U,n)$ parameter space, with $J_H/U=0.4$ and $V=0$. We see $p$-wave SC is favored for all lower values of $U$. In particular, it exists right at the type-II vHS (where $n=0.55$), while it would be unfavorable in the case of a type-I vHS. We find the critical scale $\La_c$ for SC increases with increasing $U$, and is of the order of $0.1\sim 1$ meV near the phase boundary. This indicates, unfortunately, that the SC $T_c$ in BiH may not be high. 

\section{Topology in the spin-polarized chiral $p$-wave SC state}\label{sec:chiral}

\begin{figure}
	\includegraphics[width=0.85\columnwidth]{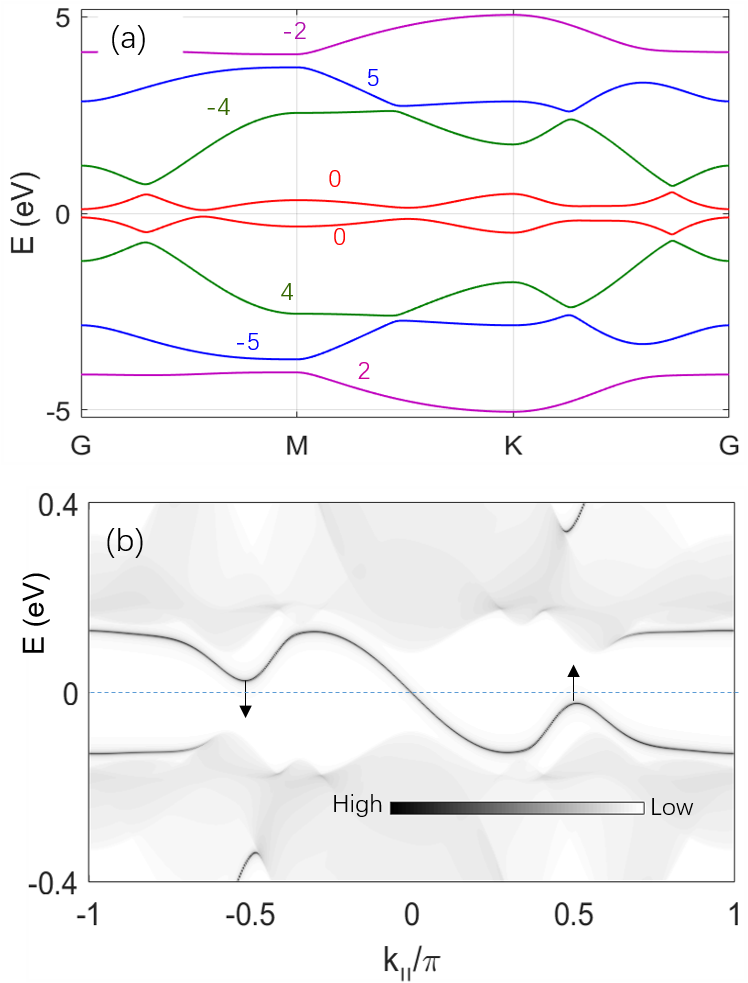}
	\caption{(a) Band dispersion along high-symmetry cuts for the spin-up BdG Hamiltonian with $\Del_0=0.2$ eV. The number near the line indicates the Chern number. To reveal the band splitting better, the green-lined bands are artificially shrinked slightly toward their own center of mass. (b) The corresponding edge spectral function $A(k_{||},E)$ (gray scale) on a zigzag edge. Here $k_{||}$ is the edge momentum and $E$ is the energy. The dashed line highlights the Fermi level, and the arrows show the movement of the ripples in the edge-state dispersion relative to the Fermi level if $\Del_0$ is reduced.}\label{fig:edge}
\end{figure}

The FRG-derived pairing function describes equal-spin pairing, and is fully polarized in each of $\phi(\v k)=i\tau_2(\si_1\pm i\si_2)i\si_2$. Here we discuss the spin-up pairing, $\phi(\v k)=i\tau_2 s_3 (\si_1+i\si_2)i\si_2 \ra i\tau_2 s_3\ua\ua$. Since $H_0$ is spin-diagonal, we solve the Bogoliubov-de Gennes (BdG) Hamiltonian in the spin-up sector,
\eqa h_{\rm BdG}(\v k)=\left(\begin{array}{cc} \eps_\v k & \Del_\v k\\ \Del_\v k^\dag & -\eps_{-\v k}^*\end{array}\right),\eea
where $\eps_\v k$ is the the normal state dispersion (a $4\times 4$ matrix in the orbital-sublattice basis) obtained from $H_0$, and $\Del_\v k=i\Del_0 \tau_2 s_3$, with a gap amplitude $\Del_0$. Note that by fixing the spin, $h_{\rm BdG}(\v k)$ is not invariant under time-reversal (TR) because of the $\tau_2$-term in $H_0$. The above BdG Hamiltonian results in eight non-degenerate bands. To gain insight into the SC state, we calculate the Chern number $C$ on each BdG band,
\eqa C_n = \frac{1}{2\pi}\int d^2 \v k (\nabla_\v k \times \v A_\v k^{(n)})_z, \eea
where the integration is over the Brillouin zone, and $A_\v k^{(n)}$ is the Berry connection on the $n$-th band,
\eqa \v A_\v k^{(n)} = -i\<\v kn|\nabla_\v k |\v kn\>, \eea
where $|\v kn\>$ is a BdG eigenstate, forming a smooth fibre bundle on the momentum manifold.
The result is indicated on each band in Fig.\ref{fig:edge}(a). Interestingly, $C=0$ for the two BdG bands near the Fermi energy (red), and high numbers are found for the other bands. However, the total Chern number is $\pm 1$ below/above the Fermi energy, showing the chiral SC state is still topologically nontrivial. Counting from the band bottom up to the Fermi level, we expect one chiral edge mode near the Fermi energy (apart from other edge modes more distant to the Fermi level). We argue that this mode, denoted as $E_{1}$, must cross the Fermi energy. On one hand, we may take the BdG bands as artificial ones for conventional fermions. We can increase the artificial chemical potential further into the conduction bands. Since the lowest conduction band is trivial, no further edge modes occur until the conduction band with $C=-4$ is reached. On the other hand, by ph symmetry of the BdG hamiltonian, $E_1$ must enter the conduction band in energy. Taken together, $E_1$ must cross the Fermi energy. This is verified in Fig.\ref{fig:edge}(b), showing the edge spectral function $A(\v k_{||},E)$, related to the electron part of the retarded Green's function, on a zigzag edge. Here $k_{||}$ is the edge momentum and $E$ the energy. We clearly see in-gap states. There are two ripples in the edge dispersion, which move along the arrows (even to the opposite side of the Fermi level) if $\Del_0$ is reduced, but this does not change the net chirality on the edge-state dispersion. We also find in-gap states on an arm-chair edge (not shown). Taken together, we find the spin-polarized state is a strong topological superconductor. The in-gap edge states are Majorana in nature, in view of the equal-spin pairing.

\section{Helical $p$-wave SC}\label{sec:helical}
The two types of pairing in $\phi(\v k)=i\tau_2 s_3(\si_1\pm i\si_2)i\si_2$, or in simpler terms, $i\tau_2 s_3(\ua\ua, \da\da)$, are block diagonal in the spin basis. Since $H_0$ is also spin-diagonal, the two spin-sectors are decoupled even in the SC state. It is interesting to ask how a TR-invariant helical SC could arise. We observe that the combination $\ua\ua + e^{i\th}\da\da=e^{i\th/2}(\ua\ua e^{-i\th/2}+\da\da e^{i\th/2})$ is helical for any phase difference $\th$, up to a global phase factor $e^{i\th/2}$ that can be absorbed by a trivial U(1) gauge transform. We need a mechanism to have equal amplitude of pairing in the above form, and we ask whether the relative phase angle could be arbitrary.

Since BiH is not mirror-symmetric about the central plane, Rashba SOC exits in principle. We model this effect as
\eqa H_{\rm SOC} = -i r\sum_{i\v b} \psi_i^\dag (b_x\si_2-b_y\si_1)\psi_{i+\v b},\eea
where $r$ is the Rashba parameter, and $\v b=(b_x,b_y,0)$ denotes a first-neighbor bond. This term breaks inversion symmetry in the effective 2d model, and could mix up the chiralities. We combine $H_{\rm SOC}$ and $H_0$ and re-perform SM-FRG. However, we find that the Rashba term, up to $r=0.08$ eV, does not break the degeneracy between the two equal-spin pairing functions. Since the applied FRG only addresses the normal state instability, it does not know whether the two degenerate spin sectors would recombine in the ordered state. For the latter purpose, we need a mean field theory
\eqa H_{\rm MF} = H_0+H_{\rm SOC} + \frac{1}{2}\sum_{\v k\si} (\Del_\si  \psi_{\v k\si}^\dag i\tau_2 s_3\psi_{-\v k\si}^{\dag,t} + {\rm h.c.}),\eea
where $\Del_\si$ is the uniform order parameter in the spin-$\si$ sector, determined self-consistently by
\eqa \Del_\si = \frac{V}{N}\sum_\v k \left\< \psi_{-\v k\si}^t i\tau_2 s_3 \psi_{\v k\si}\right\>, \label{eq:bcs}\eea
where $V$ may be taken as the 2PI part of the FRG-derived pair interaction,\cite{SM} $N$ is the number of unitcells, and $\v k$ runs in the reduced Brillouine zone. The mean field calculation at the temperature $T= 10^{-5}$ eV shows that with a finite Rashba coupling, say $r=0.03$ eV, the order parameters converge to the configuration
\eqa \Del_\da = \Del_\ua e^{i n \pi /3}, \ \ \ n=1, 2, \cdots, 6,\eea
where $n$ depends on the initial condition. Therefore, the helical SC state exists deep in the ordered state. The relative phase $\th$ is not unique, but is also not arbitrary. There is a discrete six-fold degeneracy in the helical states. This actually explains why FRG does not see the helical combination of the two equal-spin pairings (for small $r$), as follows. By the six-fold degeneracy and the symmetry of the system, we can cook up a Landau free energy in the form
\eqa f = &&\sum_\si (\al |\Del_\si|^2 + \bt |\Del_\si|^4) + \bt' |\Del_\ua|^2 \Del_\da|^2 \nn
  &&+ \ga [(\Del_\ua^*\Del_\da)^6 + {\rm c.c.}] + \cdots.\eea
The above helical states are favored by $\bt'<0$ and $\ga < 0$, but the $\ga$-term is a 12-th order term and should play no role in locking up the phase between $\Del_\ua$ and $\Del_\da$ near the transition temperature.

\begin{figure}
	\includegraphics[width=0.9\columnwidth,clip]{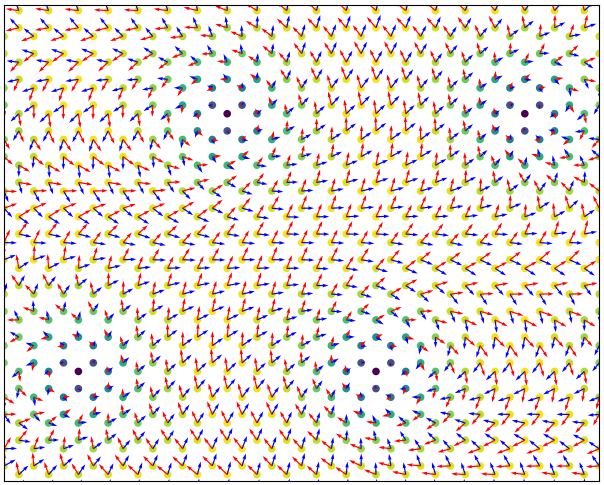}
	\caption{ Mean field solution of the triangle vortex lattice. The view field contains four vortices. The arrows show $(\Re \Del_{i\si},\Im \Del_{i\si})$ for $\si = \ua$ (blue) and $\si=\da$ (red), and the geometric mean $\Del_i=\sqrt{\sum_\si|\Del_{i\si}|^2}$ is encoded by the colored spots. }\label{fig:vortex}
\end{figure}

It is also interesting to ask whether a particular helical state is stable in a nonuniform state, such as in the vortex state, where phase winding is enforced by the magnetic field. To answer this question, we need to rewrite the mean field Hamiltonian in the real space,
\eqa H_{\rm MF}=&&\sum_{i\v b}  \psi^\dag_{i} ~t_\v b e^{-i\v A\cdot\v b} ~\psi_{i+\v b} -\sum_{i} \psi_i^\dag (\mu + \la\tau_2\si_3)\psi_i\nn
&&-i r\sum_{i\v b} \psi_i^\dag (b_x\si_2-b_y\si_1) e^{-i\v A\cdot \v b}\psi_{i+\v b}\nn
&&+\sum_{i\si}(\Del_{i\si}\psi_i^\dag i\tau_2 s_i \psi_{i\si}^{\dag,t}+{\rm h.c.}),\eea
where $\v A$ is the vector potential, and $s_i=\pm 1$ on A/B sublattice, taking the role of $s_3$ in the pairing function. The order parameter is now determined by
\eqa \Del_{i\si}=V\<\psi_{i\si}^\dag i\tau_2 s_i \psi_{i\si}^{\dag,t}\>.\eea
We performed mean field calculation of the vortex lattice. In Fig.\ref{fig:vortex} we show the geometric mean $\Del_i =\sqrt{\sum_\si |\Del_{i\si}|^2}$ (colored spots) and the phase winding (arrows) of the components $\Del_{i\ua}$ (blue) and $\Del_{i\da}$ (red) on the lattice. (We have tuned the pair interaction $V$ for the illustration of the vortex lattice as shown.) Depending on the intial condition, the solution converges at $\Del_{i\da}=\Del_{i\ua} e^{i n\pi/3}$, with $n=1$ in Fig.\ref{fig:vortex}, everywhere on the vortex lattice, showing the local robustness of the helical SC against the perturbation of the magnetic field.

\section{Summary}\label{sec:summary}

To conclude, we find triplet $p$-wave pairing is very likely in BiH. By atomic SOC, the $p$-wave pairing is between equal-spin electrons, and is chiral and two-fold degenerate. The in-gap edge states arise unusually since the BdG bands near the Fermi energy are topologically trivial. The Rashba SOC drives the chiral $p$-wave into helical $p$-wave only deep in the ordered state. The helical $p$-wave is six-fold degenerate. The relative phase angle in a helical state is shown to be stable in the vortex state.

In real materials, there may be domain walls separating helical states with different internal phase angles between the two order parameters. The consequence of such domain walls, and in particular the corner to three or more domain walls, is an interesting topic to be explored further.

\acknowledgements{QHW thanks Naoto Nagaosa and Yu-Xin Zhao for fruitful discussions. The project is supported by the National Key Research and Development Program of China (under grant No. 2016YFA0300401) and the National Natural Science Foundation of China (under grant No. 11574134 and No. 11604168).}


\begin{references}

\bibitem{Ivanov} D. A. Ivanov, Phys. Rev. Lett. {\bf 86}, 268 (2001).

\bibitem{Read} N. Read and D. Green, Phys. Rev. B. {\bf 61}, 10267 (2000).

\bibitem{Nayak} C. Nayak, S. H. Simon, A. Stern, M. Freedman, and S. Das Sarma, Rev. Mod. Phys. {\bf 80}, 1083 (2008).

\bibitem{wangyao} Y. Wang, J.-G. Liu, W.-S. Wang, and Q.-H. Wang, Phys. Rev. B. {\bf 97}, 174513 (2018).

\bibitem{frg3} Y.-Y. Xiang, W.-S. Wang, Q.-H. Wang, and D.-H. Lee, Phys. Rev. B. {\bf 86}, 024523 (2012).

\bibitem{frg7} Y. Yang, W.-S. Wang, Y.-Y. Xiang, Z.-Z. Li, and Q.-H. Wang, Phys. Rev. B {\bf 88}, 094519 (2013).

\bibitem{yaohong} H. Yao and F. Yang, Phys. Rev. B. {\bf 92}, 035132 (2015).

\bibitem{Song} Z. Song, C.-C. Liu, J. Yang, J. Han, B. Fu, Y. Yang, Q. Niu, J. Lu, and Y. G. Yao, NPG Asia Materials. {\bf 6}, e147 (2014).

\bibitem{Liu} C.-C. Liu, S. Guan, Z. Song, S. A. Yang, J. Yang, and Y. G. Yao, Phys. Rev. B. {\bf 90}, 085431 (2014).

\bibitem{Yang} F. Yang, C.-C. Liu, Y.-Z. Zhang, Ygui Yao, and Dung-Hai Lee, Phys. Rev. B. {\bf 91}, 134514 (2015).

\bibitem{frg1} Q.-H. Wang, C. Platt, Y. Yang, C. Honerkamp, F. C. Zhang, W. Hanke, T. M. Rice, and R. Thomale, EPL {\bf 104}, 17013 (2013).

\bibitem{frg2} W.-S. Wang, Y.-Y. Xiang, Q.-H. Wang, F. Wang, F. Yang, and D.-H. Lee, Phys. Rev. B. {\bf 85}, 035414 (2012).

\bibitem{frg4} W.-S. Wang, Z.-Z. Li, Y.-Y. Xiang, and Q.-H. Wang, Phys. Rev. B. {\bf 87}, 115135 (2013).

\bibitem{frg5} Y.-Y. Xiang, F. Wang, D. Wang, Q.-H. Wang, and D.-H. Lee, Phys. Rev. B {\bf 86}, 134508 (2012).

\bibitem{frg6} Y.-Y. Xiang, Y. Yang, W.-S. Wang, Z.-Z. Li, and Q.-H. Wang, Phys. Rev. B {\bf 88}, 104516 (2013).

\bibitem{frg8} W.-S. Wang, M. Gao, Y. Yang, Y.-Y. Xiang, and Q.-H. Wang, Phys. Rev. B {\bf 95}, 144507 (2017).

\bibitem{frg9} Y.-C. Liu, W.-S. Wang, F.-C. Zhang, and Q.-H. Wang, Phys. Rev. B {\bf 97}, 224522 (2018).

\bibitem{GFZhang} G.-F. Zhang, Y. Li, and C. Wu, Phys. Rev. B {\bf 90}, 075114 (2014).

\bibitem{Slater} J. C. Slater and G. F. Koster, Phys. Rev. {\bf 94}, 1498 (1954).

\bibitem{Wetterich} C. Wetterich, Phys. Lett. B. {\bf 301}, 90 (1993).

\bibitem{note}  In two dimension, a continuous symmetry can not be broken at finite temperature according to the Mermin-Wagner theorem. In this case $\La_c$ is understood as the crossover temperature below which long-range correlations are to be developed.

\bibitem{SM} Supplemental Materials at [......], describing technical details of SM-FRG and the FRG-derived mean field theory.



\end{references}
\end{document}